\documentclass[a4paper,UKenglish,cleveref, autoref]{lipics-v2016}

\usepackage{graphics}
\usepackage{tikz}
\usetikzlibrary{positioning}
\usepackage{graphicx}
\usepackage{algorithm}
\usepackage{multicol}
\usepackage{amsmath}
\usepackage{amssymb}
\usepackage{bm}
\newcommand{\remove}[1]{}
\newcommand{\h}{\hspace*{0.2in}}

\title{Parallel and Distributed Algorithms for the housing allocation Problem\footnote{This work was partially supported by NSF CSR-1563544,CNS-1812349, and Cullen Trust Professorship}} 

\titlerunning{Parallel and Distributed Algorithms for housing allocation and Related Problems}

\author[1]{Xiong Zheng}
\author[2]{Vijay K. Garg}
\affil[1]{Electrical and Computer Engineering Department, University of Texas at Austin, USA\\
  \texttt{zhengxiongtym@utexas.edu}}
\affil[2]{Electrical and Computer Engineering Department, University of Texas at Austin, USA\\
  \texttt{garg@ece.utexas.edu}}
\authorrunning{J.\,Q. Open and J.\,R. Access}

\authorrunning{X. \, Zheng and V. \, K. Garg}

\Copyright{Xiong Zheng and Vijay Garg}


\EventEditors{John Q. Open and Joan R. Access}
\EventNoEds{2}
\EventLongTitle{42nd Conference on Very Important Topics (CVIT 2016)}
\EventShortTitle{CVIT 2016}
\EventAcronym{CVIT}
\EventYear{2016}
\EventDate{December 24--27, 2016}
\EventLocation{Little Whinging, United Kingdom}
\EventLogo{}
\SeriesVolume{42}
\ArticleNo{23}

\begin{document}

\maketitle

\begin{abstract}
We propose parallel and distributed algorithms for the \textit{housing allocation} problem. In this problem, there is a set of agents and a set of houses. Each agent has a strict preference list for a subset of houses. We need to find a matching for agents to houses such that some criterion is optimized. One such criterion which has attracted much attention is \textit{Pareto Optimality}. A matching is Pareto optimal if no coalition of agents can be strictly better off by exchanging houses among themselves. We also study the \textit{housing market} problem, a variant of the housing allocation problem, where each agent initially owns a house. In addition to Pareto optimality, we are also interested in finding the \textit{core} of a housing market. A matching is in the \textit{core} if there is no coalition of agents that can be better off by breaking away from other agents and switching houses only among themselves in the initial allocation.

In the first part of this work, we show that computing a Pareto optimal matching of a house allocation is in {\bf CC} and computing the core of a housing market is {\bf CC}-hard, where {\bf CC} is the class of problems logspace reducible to the comparator circuit value problem.   
Given a matching of agents to houses, we show that verifying whether it is Pareto optimal is in {\bf NC}. We also show that verifying whether it is in the core can be done in {\bf NC}. We then give an algorithm to show that computing a maximum cardinality \textit{Pareto} optimal matching for the housing allocation problem is in {\bf RNC}\textsuperscript{2} and quasi-{\bf NC}\textsuperscript{2}.

In the second part of this work, we present a distributed version of the top trading cycle algorithm for finding the \textit{core} of a housing market. To that end, we first present two algorithms for finding all the disjoint cycles in a functional graph. The first algorithm is a \textit{Las Vegas} algorithm which terminates in $O(\log l)$ rounds with high probability, where $l$ is the length of the longest cycle. The second algorithm is a deterministic algorithm which terminates in $O(\log^* n \log l)$ rounds, where $n$ is the number of nodes in the graph. Both algorithms work in the synchronous distributed model and use messages of size $O(\log n)$. By applying these two algorithms for finding cycles in a functional graph, we give the distributed top trading cycle algorithm which terminates in $O(n)$ rounds and requires $O(n^2)$ messages. 
\end{abstract}

\section{Introduction}
Matching is a fundamental problem in computer science with numerous applications. The \textit{housing allocation}  problem \cite{hylland1979efficient, zhou1990conjecture, abdulkadirouglu1998random} is an instance of matching problem with one-sided preferences. In this problem, we need to allocate a set $H$ of houses among a set $A$ of agents and monetary compensations are not allowed. Each agent $a_i \in A$ ranks in order of preference a subset of $H$ (the acceptable houses for $a_i$). The variant in which there is an initial endowment of houses to agents is known as the \textit{housing market} problem \cite{shapley1974cores, roth1977weak,roth1982incentive}. For both the housing market and the housing allocation problem, we need to construct a matching of agents to houses such that it is optimal with respect to some criterion. One criterion usually considered is \textit{Pareto Optimality} \cite{abdulkadirouglu1998random, abraham2004pareto, shapley1974cores}. A matching $M$ is Pareto optimal if there is no other matching $M'$ such that no agent strictly prefer $M$ to $M'$, and at least one agent strictly prefer $M'$ to $M$. For example, a matching $M$ is not Pareto optimal if a group of agents could improve by exchanging the houses that they are assigned to in $M$. 

Possible applications of the housing allocation problem and the housing market problem include: assigning virtual machines to servers in cloud computers, allocating graduates to trainee positions, professors to offices, and students to roommates. Yuan \cite{yuan1996residence} also describes a large-scale application of housing allocation in the allocation of families to government-subsidized housing in China. Also, the paper \cite{maggs2015algorithmic} describes applications of algorithms for the stable marriage problem  for mapping clients to server clusters in a content delivery network in Akamai. When only one side preference is considered, housing allocation algorithms can be applied. 

For the housing allocation problem, there is a simple greedy algorithm, known as the \textit{serial dictatorship mechanism} \cite{abdulkadirouglu1998random} to compute a Pareto optimal matching. 
The \textit{serial dictatorship mechanism} works as follows. Arbitrate a total ordering on all the agents. Let all agents pick their top choice of the remaining houses one by one following the total order. This algorithm is sequential and takes $O(n^2)$ computation steps. Also, it does not necessarily give a maximum cardinality \textit{Pareto} optimal matching. 
The paper \cite{abraham2004pareto} studies the problem of finding a maximum cardinality Pareto optimal matching for the housing allocation problem in the sequential setting. Their algorithm first computes a maximum cardinality matching of the bipartite graph formed by agents and houses and then improves the matching to be Pareto optimal. Their algorithm runs in $O(\sqrt{n}m)$ sequential time, where $n$ is the number of agents plus the number of houses, and $m$ is the number of edges of the agent-house bipartite graph. They also show that any improvement to the complexity of their algorithm would imply an improved algorithm for finding a maximum matching in a bipartite graph.

For the housing market problem, Shapley and Scarf \cite{shapley1974cores} prove that there exists at least one matching in the core of any housing market and present the well-known \textit{top trading cycle} (TTC) mechanism, which they attribute to David Gale. This mechanism works by repeatedly finding the top preference cycles and exchanging houses along those cycles. It takes $O(n^2)$ sequential steps. Ma \cite{ma1994strategy} shows that the TTC mechanism is the only individually rational, Pareto-efficient, and strategy-proof mechanism. Roth and Postlewaite \cite{roth1977weak} show that there is exactly one core for each housing market instance. Note that the matching obtained by the TTC mechanism is not only a \textit{Pareto} optimal maching, but also the unique \textit{core}. 

The parallel complexity of both these problems has not been studied in the literature. The housing allocation problem is a variant of the stable marriage problem with only one sided preferences. The decision version of the stable marriage problem, i.e, given a pair of man and woman, to decide whether they are matched in the man-optimal stable matching, is {\bf CC}-complete \cite{mayr1992complexity}. The {\bf CC} class \cite{mayr1992complexity} is the set of problems logspace reducible to the comparator circuit value problem. Currently, there are no efficient parallel algorithms for this class of problems. It is conjectured that {\bf CC} is incomparable with {\bf NC} \cite{cook2014complexity, mayr1992complexity}, the class of problems computable in polylog parallel time. In this work, we show that finding the core of a housing market is {\bf CC}-hard, which can be taken as evidence that this problem is not parallelizable.  Although finding the core is hard, we show that given a matching, it can be verified in {\bf NC} whether it is the core. On the other hand, finding a Pareto optimal matching is easier than finding the core. We show that finding a maximum cardinality Pareto optimal matching can be done in {\bf RNC}\textsuperscript{2} and quasi-{\bf NC}\textsuperscript{2}, where {\bf RNC}\textsuperscript{2} represents the problems which have uniform circuits of polynomial size and $O(\log^2 n)$ depth and quasi-{\bf NC}\textsuperscript{2} represents the problems which have uniform circuits of quasi-polynomial size $n^{O(\log n)}$ , and $O(\log^2 n)$ depth.

In this paper, we also study the housing market problem in the distributed setting. Specifically, we give a symmetric distributed algorithm for the TTC mechanism. By symmetric, we mean that each agent performs the same role. 

In summary, this paper makes the following contributions:
\begin{itemize}
\item We prove that computing the core of a housing market is {\bf CC}-hard, by giving a logspace reduction from the lexicographically first maximal matching problem, which is a {\bf CC}-complete problem, to the housing market problem.


\item We show that computing a maximum cadinality Pareto optimal matching for the housing allocation problem is in {\bf RNC}\textsuperscript{2} and quasi-{\bf NC}\textsuperscript{2}. 

\item We give a symmetric distributed TTC algorithm for computing the core of a housing market, which runs in $O(n)$ rounds and require $O(n^2)$ messages. 

\end{itemize}


The paper is organized as follows. Section 2 gives preliminaries for the housing allocation and the housing market problem.
Section 3 studies the parallel complexity of the housing market problem. 
Section 4 presents a parallel algorithm for computing a maximum cardinality Pareto optimal matching for the housing allocation. 
Section 5 presents a distributed algorithm for computing the core of a housing market.
Finally, section 6 present the conclusions and future work.

\section{Preliminaries} 
The \textit{housing allocation} problem deals with assigning indivisible houses
to agents who have preferences over these houses. In general, a housing allocation instance $(A,H,P)$ consists of \\
\h (1) a set of agents $A = \{a_1,a_2,...,a_n\}$,\\
\h (2) a set of indivisible houses $H=\{h_1,h_2,...,h_m$\},\\ 
\h (3) a preference profile $P=\{\prec_{a_1}, \prec_{a_2},...,\prec_{a_n}\}$, where $\prec_{a_i}$ defines a strict preference of \\
\h\h agent $a_i$ on a subset of houses. \\
 We restrict our attention to strict preference profiles where each agent defines a strict total order over a subset of houses. Let $N(i)$ denote the subset of acceptable houses for agent $i$. The goal of the housing allocation problem is to find a \textit{Pareto} optimal matching of agents to houses. For a matching $\mu$ and an agent $i$, let $\mu(i)$ denotes the house matched to agent $i$. We use $h \prec_i h'$ to denote that agent $i$ prefers house $h$ to house $h'$. For two matchings $\mu$ and $\nu$, $\mu \prec_i \nu$ denotes that agent $i$ prefer $\mu(i)$ to $\nu(i)$.  
The definitions of \textit{Pareto Domination} and \textit{Pareto Optimality} \cite{shapley1974cores} are given as below.

\begin{definition} (Pareto Domination). Suppose $\mu, \nu$ are matchings. Then $\mu$ 
Pareto dominates $\nu$ if and only if

(1) $\mu \preceq_i \nu$ for all $i \in A$,

(2) $\mu \prec_j \nu$ for some $j \in A$.
\end{definition}

We use $\mu \prec \nu$ to denote that matching $\mu$ Pareto dominates matching $\nu$. 

\begin{definition} (Pareto Optimality). Suppose $\mu$ is a matching. Then $\mu$ is Pareto
optimal if and only if it is not Pareto dominated by any other matching $\nu$.  
\end{definition}

The \textit{housing market} problem is a variant of the housing allocation problem, where there is an initial endowment of houses to agents and we have the same number of agents and houses. Let $\mu_0$ be a matching denoting the initial endowment of houses to agents. Let $(A,H,P,\mu_0)$ denote an instance of a housing market. In the housing market problem, in addition to Pareto optimality, we also want a matching to be individually rational \cite{abdulkadirouglu1999house} defined as follows. 

\begin{definition} (Individually Rational). Suppose $\mu$ is a matching of agents to houses in a housing market. Then $\mu$ is individually rational if
$\mu(a) \preceq_a \mu_0(a)$ for all $a \in A$. 
\end{definition}

Individual rationality means an agent is willing to give up its initially assigned house only when it can get a better house. 
To define the \textit{core} of a housing market, let us first define the concept of \textit{coalition}. Informally, given a matching $\mu$, a coalition w.r.t $\mu$ is a set of agents $A' \in A$ such that, by only switching houses within themselves, each agent in $A'$ can get a house at least as good as its house in $\mu$ and at least one agent gets a strictly better house. 

\begin{definition} (Coalition). Given a housing market $(A,H,P, \mu_0)$ and a matching $\mu$, a set of agents $A' \subseteq A$ form a coalition w.r.t $\mu$ if there exists a matching $\nu$ such that 

(1) $\nu(a) \in \{\mu_0(b)~ | ~b \in A'\}, ~ \forall a \in A'$ 

(2) $\nu(a) \preceq_a \mu(a) ~\forall a \in A'$

(3) $\exists ~a \in A' ~\text{such that} ~ \nu(a) \prec_a \mu(a)$
\end{definition}
Condition (1) says that to get matching $\nu$ from $\mu_0$, the agents in $A'$ only switch houses within themselves. Condition (2) means that in matching $\nu$ each agent in $A'$ is matched to a house at least as good as the house it gets matched to in $\mu$. Condition (3) means that at least one agent is matched to a better house in matching $\nu$. The \textit{core} \cite{shapley1974cores} of a housing market is defined as follows. 

\begin{definition} (Housing Market Core). The core of a housing market problem is a set of matchings $M$ such that matching $\mu \in M$ if and
only if there does not exist any coalition $A'$ w.r.t $\mu$. 
\end{definition}

Essentially, a matching is in the core of a housing market if there does not exist a set of agents such that they can match to better houses by breaking away from other agents and exchanging houses within themselves. An individually rational and Pareto optimal matching is not necessarily a core matching, whereas a core matching must be individually rational and Pareto optimal. An example to illustrate the difference between a core matching and an individually rational Pareto optimal matching is given in Fig. \ref{fig:core-individual-pareto}.

\begin{figure}[!h] 
\centering
\begin{tikzpicture}[
roundnode/.style={circle, draw=green!60, very thick, minimum size=7mm},
covernode/.style={circle, draw=green!60, fill=green!60, very thick, minimum size=7mm},
]

\node			(a1)       					    {$a_1: h_2, h_3, h_1$};
\node			(a2)        
[below=0.05in of a1] 		{$a_2:h_1,h_3,h_2$};
\node			(a3)			[below=0.05in of a2]
{$a_3:h_1,h_2,h_3$};
\node (t1) [below=0.05in of a3] {Preference Profile};

\node (i1) [right=0.5in of a1]  {$a_1:h_1$};
\node (i2) [right=0.5in of a2] {$a_2:h_2$} ;

\node (i3) [right=0.5in of a3] {$a_3:h_3$} ;
\node (t2) [below=0.05in of i3]  {Initial Endowment};

\node (p1) [right=1in of i1] {$a_1:h_2$} ;

\node (p2) [right=1in of i2] {$a_2:h_3$} ;

\node (p3) [right=1in of i3] {$a_3:h_1$} ;
\node (t3) [below=0.05in of p3] {$M_1$};

\node (c1) [right=1in of p1] {$a_1:h_2$} ;

\node (c2) [right=1in of p2] {$a_2:h_1$} ;

\node (c3) [right=1in of p3] {$a_3:h_3$} ;
\node (t3) [below=0.05in of c3]
{$M_2$};

\end{tikzpicture}
\caption{An Example\label{fig:core-individual-pareto}}
\end{figure}

It is easy to see that both $M_1$ and $M_2$ are individually rational and Pareto optimal matchings. $M_2$ is the core matching but $M_1$ is not. In $M_1$, agents $a_1$ and $a_2$ can form a coalition within themselves and swap houses. Suppose $a_1$ and $a_2$ break away from other agents and switch houses with each other. Then, $a_1$ gets house $h_2$, which is the same as the house it gets in $M_1$, and $a_2$ gets house $h_1$, which is strictly better than the house it gets in $M_1$. Thus, $M_1$ is not the core. On the other hand, in $M_2$, there does not exist two agents such that at least one will be strictly better off by forming a coalition and swapping houses among themselves. 

The following result is well-known. 
\begin{lemma} [\cite{roth1977weak}] 
There is exactly one unique matching in the core of a housing market instance. 
\end{lemma}

Since the core of a housing market has one unique matching, we use the core to mean this unique matching henceforth. The TTC algorithm given by Shapley and Scarf \cite{shapley1974cores} computes the unique core of a housing market. This algorithm works in stages. At each stage, it has the following steps: \\
\h Step 1. Construct the \textit{top choice} directed graph $G_t=(A,E)$ on the set of agents as follows. Add an arc from agent $i \in A$ to agent $j \in A$ if $j$ holds the current top house of $i$.\\
\h Step 2. Since each node has exactly one outgoing edge in $G_t$, there must be at least one cycle, which could be a self-loop. All cycles are node disjoint. Find all the cycles in the top trading graph and implement the trade indicated by the cycles, i.e, each agent which is in any cycle gets its current top house. \\ 
\h Step 3. Remove all agents which get their current top houses and remove all houses which are assigned to some agent from the preference list of remaining agents. 

The above steps are repeated until each agent is assigned a house. At each stage, at least one agent is assigned a final house. Thus, this algorithm takes $O(n)$ stages in the worse case and needs $O(n^2)$ computational steps.

\section{Parallel Algorithms for Housing allocation and Housing Market} 
In this section, we study the parallel complexity of the housing allocation and housing market problem. The parallel computation model we use here is the CREW PRAM model \cite{karp1988survey}. First, we show that computing a Pareto optimal matching in a housing allocation is {\bf CC} by reducing this problem to the lexicographically first maximal matching problem (LFMM), which is a {\bf CC}-complete problem \cite{mayr1992complexity}. In the LFMM problem, we are given a graph $G(V,E, \prec)$ where $\prec$ denotes a total ordering on the edges. If $e_1 \prec e_2$, we say that $e_1$ precedes $e_2$. The total order $\prec$ allows us to regard a matching $M$ as a sequence $S_M = (e_1,e_2,...)$ of edges in ascending order, i.e, $j < k \implies e_j \prec e_k$. Given two matchings $M$ and $N$, we say $M \prec N$ if $S_M$ lexicographically precedes $S_N$. The relation $\prec$ defines a total order over all maximal matchings. The minimum element, $M_{lex}$, of this order is call the lex-first maximal matching of $G(V,E,\prec)$. We need to decide whether a given edge $e$ is in the lex-first maximal matching of the graph. Then, we show that computing the unique core of a housing market is {\bf CC}-hard, by giving a logspace reduction from the LFMM problem to the housing market problem. We say a problem is {\bf CC}-hard if every problem in {\bf CC} reduces to it. 

\begin{theorem} \label{the:pareto-cc}
Computing a Pareto optimal matching for a housing allocation is in {\bf CC}. 
\begin{proof}
We reduce the problem of computing a Pareto optimal matching to the LFMM problem. Given a housing allocation instance $(A,H,P)$, we construct an agent-house bipartite graph $G=(A \cup H,E,\prec)$ where $\prec$ denotes a total ordering of edges. There is an edge from an agent $u \in A$ to a house $v \in H$ if $v$ is acceptable to $u$. For each agent $u$, let $r_u: H \rightarrow [|N(u)|]$ denote its rank function, i.e, $r_u(v)$ denote the rank of house $v$ at agent $u$'s preference list. We assign an arbitrary unique ordering to all the agents, i.e, a one-to-one function $f:A \rightarrow [n]$, where $n= |A|$. For each edge $(u,v)$, we associate the tuple $<f(u),r_u(v)>$ with it. We define the total order $\prec$ on edges as the lexicographical ordering of the tuples associated with them. Clearly, $\prec$ defines a total ordering on all edges. We claim that the lex-first maximal matching $M_{lex}$ of this graph corresponds to a Pareto optimal matching. Suppose not, then there exists another maximal matching $M'$ which dominates $M$. There must exist an agent $u \in A$ such that $u$ prefers $M'(u)$ to $M(u)$. We have that $r_u(M'(u)) < r_u(M(u))$, which means that $(u,M'(u)) \prec (u,M(u))$, contradicting the fact that $M$ is the lex-first maximal matching. 
\end{proof}
\end{theorem}

\begin{corollary}
There is a $\Tilde{O}(\sqrt{|E|})$ time parallel algorithm which uses $O(|E|)$ processors to compute a Pareto optimal matching, where is $|E|$ is the number of acceptable agent-house pairs. 
\begin{proof}
Follows from the fact that there is a $\Tilde{O}(\sqrt{|E|})$ time parallel algorithm for the LFMM problem \cite{mayr1992complexity}. 
\end{proof}
\end{corollary}

\begin{corollary}
There is a $O(\sqrt{n})$ round distributed algorithm in the congest clique model for computing a Pareto optimal matching, where is $n$ is the number of agents and houses. 
\begin{proof}
The paper \cite{amira2010distributed} gives a $O(\sqrt{n})$ distributed algorithm for the weighted stable marriage problem. The LFMM problem is simply a subcase of the weighted stable marriage problem. Thus, the same algorithm can be applied here.
\end{proof}
\end{corollary}

\begin{remark}
In the housing market problem, an individual rational and Pareto optimal matching must be a perfect matching of houses to agents. Thus, the reduction given in Theorem \ref{the:pareto-cc} cannot be applied. Instead, the problem of computing a individual rational and Pareto optimal matching can be reduced to the problem of lex-first perfect matching. Unfortunately, the complexity of this problem is unknown. 
\end{remark}

In a housing market $(A,H,P,\mu_0)$, the weighted agent-house bipartite graph $G=(A \cup H, E, w)$ is defined as follows. There is an edge between agent $a_i \in A$ and house $h_j \in H$ if either $h_j = \mu_0(a_i)$ or $a_i$ prefers $h_j$ to $I(a_i)$. The weight of the edge is defined as the the rank of $h_j$ in $a_i$'s preference list. To compute an individual rational and Pareto optimal matching for a housing market, we first observe the following lemma from \cite{david2013algorithmics}. 

\begin{lemma} [\cite{david2013algorithmics}] \label{lem:minimum-pareto}
A minimum weight perfect matching of the weighted agent-house bipartite graph is an individual rational and Pareto optimal matching. 
\begin{proof}
Let $u$ be a minimum perfecting matching of the agent-house bipartite graph. Suppose $u$ is not Pareto optimal. Then there must be another perfect matching $v$ such that $v$ Pareto dominates $u$. By the definition of Pareto domination, we can easily argue that $v$ has smaller weight than $u$, contradiction. 
\end{proof}
\end{lemma}

Combining with the results from \cite{fenner2016bipartite} and \cite{mulmuley1987matching}, we have the following result. 
\begin{theorem} [\cite{david2013algorithmics,fenner2016bipartite, mulmuley1987matching}]\label{the:RNC2}
There is a {\bf RNC}\textsuperscript{2} algorithm and a quasi-{\bf NC}\textsuperscript{2} algorithm for computing a individually rational and Pareto optimal matching of a housing market, which require $O(n^3\cdot m)$ and $n^{O(\log n)}$ parallel processors, respectively. 
\begin{proof}
By Lemma \ref{lem:minimum-pareto}, to compute an individually rational and Pareto optimal matching, we can find a minimum weight perfect matching of the agent-house bipartite graph. By the results of \cite{mulmuley1987matching} and \cite{fenner2016bipartite}, computing the minimum weight perfect matching of a bipartite graph is in {\bf RNC}\textsuperscript{2} and quasi-{\bf NC}\textsuperscript{2}. 
\end{proof}
\end{theorem}

Now we show that computing the core of a housing market is {\bf CC}-hard. 
\begin{theorem} \label{the:cc-complete}
Computing the core of a housing market is {\bf CC}-hard. 
\begin{proof}
We reduce the LFMM problem to the housing market problem. 
Let $G=(V, E, \prec)$ be an instance of a LFMM problem, where $\prec$ represents the total ordering on the edges. Let $M_{lex}$ denote the lex-first maximal matching of $G$. We construct an instance for the housing market problem as follows. For each node $v \in V$, we create an agent $a_v$ and a house $h_v$. So, we have $|A| = |H| = |V|$. Each agent $a_v$ is initially assigned house $h_v$. The preference list for each agent is constructed based on the total ordering of edges in $E$. Note that to compute the core of a housing market, the preference list of an agent below its initial assigned house is irrelevant, since the core must be individually rational. So for each agent, we only need to specify the part of the preference list above its initial assigned house. For each pair of agents $a_u$ and $a_v$, if edge $(u,v)$ exists in graph $G$, then agent $a_u$ prefers the house $h_v$ of agent $a_v$ to its own house $h_u$ . Otherwise, agent $a_u$ prefers its own house $h_u$ to $h_v$. In other words, for each edge $(u,v) \in E$, agent $a_u$ prefers house $h_v$ of agent $a_v$ to its own house $h_u$. The preference list of an agent $a_u$ is defined based on the order of edges incident to vertex $u$, i.e, agent $a_u$ prefers the house $h_v$ of agent $a_v$ to the house $h_w$ of agent $a_w$ if $(u,v) \prec (u,w)$. Since all edges are totally ordered, the preference list for each agent is strict. Fig. \ref{fig:lffm-ha} shows an example of reduction above. Clearly, the above reduction can be done in logarithmic space. 

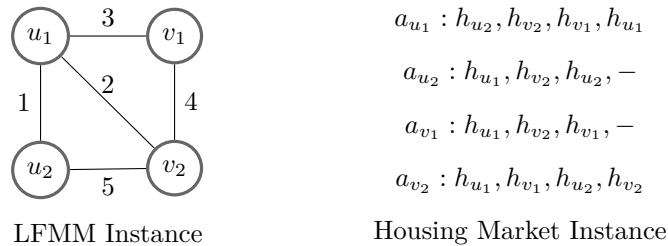
\begin{figure}[htbp] 
\centering
\begin{tikzpicture}[
roundnode/.style={circle, draw=green!60, very thick, minimum size=7mm},
covernode/.style={circle, draw=black!60, very thick,minimum size=1mm},
]

\node[covernode]			(l1)       				{$u_1$}	    ;
\node[covernode]			(l2)        [below=of l1] 		{$u_2$};

\node[covernode]			(r1)       		[right=of l1]		{$v_1$}	    ;
\node[covernode]			(r2)       			[below=of r1]	{$v_2$}	    ;

\draw[-] (l1) -- (r1);
\draw[-] (l1) -- (r2);
\draw[-] (l2) -- (r2);
\draw[-] (l1) -- (l2);
\draw[-] (r1) -- (r2);

\path (l1) -- node [auto=false,thick,above]{$3$} (r1);
\path (l1) -- node[auto=false,thick,above]{$2$} (r2);
\path (l1) -- node[auto=false,thick,left]{$1$} (l2);
\path (r1) -- node[auto=false,thick,right]{$4$} (r2);
\path (r2) -- node (p1) [below]{$5$} (l2);
\node (t1) [below=0.05in of p1] {LFMM Instance};

\node (t2) [right=2cm of t1] {Housing Market Instance};
\node			(a4)			[above =0.05in of t2]
{$a_{v_2}:h_{u_1},h_{v_1},h_{u_2},h_{v_2}$};
\node			(a3)			[above=0.05in of a4]
{$a_{v_1}:h_{u_1},h_{v_2},h_{v_1},-$};
\node			(a2)        
[above =0.05in of a3] 		{$a_{u_2}:h_{u_1},h_{v_2},h_{u_2},-$};
\node			(a1)       			[above=0.05in of a2]		    {$a_{u_1}: h_{u_2}, h_{v_2}, h_{v_1},h_{u_1}$};

\end{tikzpicture} 

\caption{Constructing a Housing Market Instance from a LFMM Instance. At stage 0, edge $(u_1,u_2)$ is added into $M_{lex}$ by the greedy algorithm and vertices $u_1$ and $u_2$ are removed from the graph.  In the TTC algorithm, the top choice graph only has one top trading cycle formed by agent $a_{u_1}$ and $a_{u_2}$. Thus, agent $a_{u_1}$ and $a_{u_2}$ switch their houses and their houses are removed from the preference list of remaining agents. At stage 1, edge $(v_1,v_2)$ is added into $M_{lex}$ by the greedy algorithm. Agents $a_{v_1}$ and $a_{v_2}$ form a top trading cycle and switch houses with each other in the TTC algorithm. }
\label{fig:lffm-ha}
\end{figure}

We claim that an edge $e=(u,v) \in E$ is in $M_{lex}$ if and only if agent $a_u$ and $a_v$ switch houses with each other in the core of the housing market instance. We say that an edge is minimum in its neighborhood if it is smaller than all its neighboring edges. Recall that the greedy algorithm for LFMM works as follows. Add each edge which is minimum in its neighborhood in the current graph into $M_{lex}$ and remove all incident edges of the two endpoints of each such edge. Repeat the above procedure until the graph is empty. Since the greedy algorithm computes the unique lex-first maximal matching of a LFMM instance and the TTC algorithm computes the unique core of a housing market,  it suffices to show that the TTC algorithm on the housing market instance simulates the greedy algorithm on $G$.  


Let $G^i=(V^i,E^i)$ denote the reduced graph at the beginning of stage $i$ of the greedy algorithm. $G^0 = G$. Let $R^i$ denote the set of edges in $E^i$ added into $M_{lex}$ by the greedy algorithm at stage $i$, i.e, the set of edges which are minimum in their neighborhoods in $G^i$. Let $M_{lex}^{i}$ denote the set of edges in $M_{lex}$ at the end of stage $i$. Let $G_t^i$ denote the top choice graph formed by remaining agents at stage $i$ of the TTC algorithm. We now show by induction on stages that an edge $(u,v)$ is added into $M_{lex}$ at stage $i$ of the greedy algorithm iff the corresponding agents $a_u$ and $a_v$ switches houses at stage $i$ of the TTC algorithm. 

Base case: stage 0. Consider an edge $e=(u,v) \in R^0$. In the housing market, two agents $a_u$ and $a_v$ correspond to this edge. Since $e$ is the minimum in its neighborhood, agent $a_u$ and agent $a_v$ are the top choice of each other. Thus, they form a top trading cycle of length 2 in $G_t^0$ and switch houses with each other in the TTC algorithm. Therefore, all edges in $R^0$ correspond to the top trading cycles in $G_t^0$.  

Induction case: assume the claim holds for stage $i$. Consider stage $i + 1$ of both algorithms. At the end of stage $i$, in the greedy algorithm, all edges incident to edges in $R^i$ are removed from the graph. In the TTC algorithm, all houses involved in the top trading cycles are removed from the preference list of remaining agents. We claim for each edge $e=(u,v) \in R^{i+1}$, the two corresponding agents $a_u$ and $a_v$ in the housing market form a top trading cycle of length 2 in $G_t^{i+1}$. Suppose not. Let $e'=(u',v') \in R^{i+1}$ be an edge such that agent $a_{u'}$ and agent $a_{v'}$ do not form a top trading cycle in $G_t^{i+1}$. We must have that either house $h_{v'}$ is not the top choice of agent $a_u$ or house $h_{u'}$ is not the top choice of agent $a_{v'}$ or both. Without loss of generality, assume house $h_{v'}$ is not the top choice of agent $a_{u'}$. We have two cases. \\
Case 1: house $h_{v'}$ is not available for agent $a_{u'}$. Then, agent $a_{v'}$ participates in a certain top trading cycle before stage $i+1$. By induction assumption, this means that there exists one edge $e_{v'}$ incident to vertex $v'$ which is added into $M_{lex}$ at a stage before $i+1$, contradicting the fact that edge $(u',v')$ exists in $E^{i+1}$. \\
Case 2: house $h_{v'}$ is available for agent $a_{u'}$ but is not the current top choice for $a_{u'}$. Then, there exists another agent $a_{w'}$ such that agent $a_{u'}$ prefers the house $h_{w'}$ of $a_{w'}$ to the house of agent $a_{v'}$. The existence of agent $a_{w'}$ indicates that it is not involved in any top trading cycle before stage $i+1$. By induction assumption, there does not exist any edge $e_{w'}$ in $M_{lex}^i$ which is incident to vertex $w'$. Thus, we have $(u',w') \in E^{i+1}$. The fact that agent $a_{u'}$ prefers the house of agent $a_{w'}$ to the house of $a_{v'}$ indicates that $(u',w') \prec (u',v')$ which contradicts the fact that $(u',v')$ is minimum in its neighborhood. 

\end{proof}
\end{theorem}

Even though we do not know any {\bf NC} algorithm for either computing an individual rational and Pareto optimal matching or computing the core of a housing market, given a matching, we can verify whether it is an individual rational and Pareto optimal matching and whether it is the core in {\bf NC}. 
\begin{theorem}
Given a matching $\mu$ of houses to agents in a housing market $(A,H,P, \mu_0)$, the following two tasks can be performed in {\bf NC}. 

1) Verifying whether $\mu$ is individual rational and Pareto optimal.
2) Verifying whether $\mu$ is the unique core. 
\begin{proof}
1). Given a matching $\mu$, to check whether it is individual rational, we just need to check whether we have $\mu \prec_{a}\mu_0$ for each agent $a \in A$. Observe that if $\mu$ is individually rational, it must be a perfect matching. Then, to verify whether it is Pareto optimal, we construct a directed graph $G=(A,E)$ as follows. There is an arc $<u,v>$ from agent $u$ to $v$ if agent $u$ prefers $\mu(v)$ to $\mu(u)$. We claim that $\mu$ is Pareto optimal iff $G$ is cycle-free. 

($\Rightarrow$). Suppose $G$ is not cycle-free. Then there exists a directed cycle in $G$. Since each arc $<u,v>$ in $G$ represents the fact that agent $u$ prefers the house of agent $v$ in matching $\mu$. Switching houses following the cycle gives better houses for all agents in the cycle, which gives a new matching $\nu$ and $\nu \prec \mu$, contradiction to the fact that $\mu$ is Pareto optimal. 

($\Leftarrow$). Suppose that $\mu$ is not Pareto optimal. Then there exists another matching $\nu$ such that $\nu \prec \mu$. The difference between $\nu$ and $\mu$ is a set of disjoint cycles. Since each agent in $\nu$ is matched to a house at least as good in $\mu$, each such cycle must be a directed cycle in $G$. 

Checking whether $G$ is cycle-free can be done in {\bf NC} by first computing the transitive closure $TC$ of $G$ and for each vertex $u$, checking whether there exists a vertex $v$ such that $TC(u,v) = 1$ and $TC(v,u) = 1$. Thus, verifying whether a given matching is individual rational and Pareto optimal can be done in {\bf NC}.

2). Given a matching $\mu$, to verify whether it is the core, we construct a different directed graph $G'(V',E')$ as follows. $V'$ represents the set of agents. There are two types of arcs: solid arcs and dashed arcs. There is a solid arc from agent $u$ to agent $v$ if $\mu(u) = \mu_0(v)$, i.e, agent $u$ is assigned the house owned by agent $v$. Hence, the solid arcs represent how agents switch houses to get matching $\mu$ from the initial matching $\mu_0$. Thus, all solid arcs form a set of disjoint directed cycles. There is a dashed arc from agent $u$ to agent $v$ if $\mu_0(v) \prec_u \mu(u)$. 

We claim that $\mu$ is the core iff there is no directed cycle which contains dashed arcs in $G$. We show that any directed cycle with at least one dashed arc represents a coalition, w.r.t $\mu$. 
Let $C$ be such a cycle in $G$. If we switch houses following the cycle $C$, i.e, for each arc $<u,v>$ in $C$, agent $u$ matches to house $\mu_0(v)$. For each solid arc $<u,v> \in C$, we have $\mu(u) = \mu_0(v)$ by the definition of solid arc. Thus, each agent with a solid outgoing arc matches to the same house as in $\mu$. For each dashed arc $<u,v> \in C$, we have $\mu_0(v) \prec_u \mu(u)$, thus each agent with a dashed outgoing arc matches to a house strictly better than its house in $\mu$. Thus, each directed cycle with at least one dashed edge represents a coalition of agents, w.r.t $\mu$. Since a matching $\mu$ is in the core iff there is no coalition with respect to $\mu$ and each directed cycle with dashed arcs represent a coalition, we get our desired claim.

To check whether there exists a directed cycle with at least one dashed arc in $G'$, we first compute the transitive closure $TC'$ of $G'$. For each dashed arc $<u,v>$, check in parallel whether $TC'(v, u) = 1$. Thus verifying whether a matching is the core can be done in {\bf NC}. 
\end{proof}
\end{theorem}

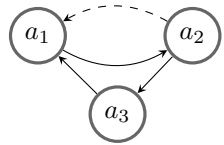
\begin{figure}[!h] 
\centering
\begin{tikzpicture}[
roundnode/.style={circle, draw=green!60, very thick, minimum size=7mm},
covernode/.style={circle, draw=black!60, very thick,minimum size=1mm},
->,>=stealth,auto,node distance=3cm
]

\node[covernode]			(a1)       					    {$a_1$};
\node[covernode]			(a2)        
[right=0.5in of a1] 		{$a_2$};
\node[covernode]			(a3)			[below right =0.5 and 0.2in of a1]
{$a_3$};

\draw[->] (a2) -- (a3);
\draw[->] (a3) -- (a1);
\path[] 
(a1) edge[bend right] node [left] {} (a2);
\path[dashed,->,every node/.style={font=\sffamily\small}] 
(a2) edge[bend right] node [left] {} (a1);
\end{tikzpicture}
\caption{To verify whether matching $M_1$ in the example given in Fig. \ref{fig:core-individual-pareto} is the core. The solid arcs represents how agents switch houses to get matching $M_1$. In $M_1$, $a_2$ prefers house $h_1$ which is the initial house of $a_1$, thus there is dashed arc from $a_2$ to $a_1$. $a_1$ and $a_2$ form a coalition w.r.t $M_1$. \label{fig:core-verification}}
\end{figure}

\subsection{A Parallel Algorithm for Maximum Pareto Optimal Matching}
In the housing allocation problem, a Pareto optimal matching does not necessarily have maximum cardinality, i.e, with maximum number of agents matched to a house. To find a maximum cardinality Pareto optimal matching, we adapt the sequential algorithm in \cite{abraham2004pareto} to be a parallel algorithm. The sequential algorithm in \cite{abraham2004pareto} has three steps. To ensure that the final matching has maximum cardinality, step 1 computes a maximum cardinality matching. After step 1, all unmatched agents are removed from consideration. At step 2, the algorithm improves the matching obtained from step 1 to be \textit{trade-in-free}. A matching $M$ is \textit{trade-in-free} if there is no (agent,house) pair $(a_i, h_j)$ such that $a_i$
is matched in $M$, $h_j$ is unmatched in $M$, and $a_i$ prefers $h_j$ to $M(a_i)$. That is, step 2 ensures that no matched agents prefers an unmatched house to its current matched house. After step 2, all unmatched houses are removed from consideration, since no matched agents prefer any of those houses to their matched houses. The final step is to improve the matching obtained from step 2 to be Pareto optimal, which is achieved by directly applying the TTC mechanism on all matched agents.

Our parallel algorithm, shown in Fig. \ref{fig:house-allocation} has only two steps. At step 1, we compute the maximum cardinality matching, which can be reduced to compute a minimum weight perfect matching of a new graph. 
Let $M'$ be the maximum cardinality matching obtained at step 1. Let $A'$ be the set of matched agents. After step 1, all the unmatched agents are removed from consideration. At step 2, we improve the matching obtained from step 1 to be Pareto optimal. In contrast to \cite{abraham2004pareto}, we do not first make our matching \textit{trade-in-free} and then \textit{Coalition-free}. Instead, we directly compute a Pareto optimal matching by computing a minimum weight perfect matching of a graph $G'$ constructed as follows. We create a set of virtual agents $B'$ to ensure the number of agents is equal to the number of houses. Add an edge with weight 0 between each virtual agent and each house. For each real agent $a_i \in A'$ and each house $h_j \in H$, add an edge between $a_i$ and $h_j$ if $h_j$ is $a_i$'s partner at the end of step 1 or $a_i$ prefers $h_j$ to its partner. The weight of edge $(a_i, h_j)$ is equal to the rank of $h_j$ in $a_i$'s preference list. 

\begin{figure}[htb] 
\centering
\fbox{\begin{minipage}[t]  {5in}
\underline{\textbf{Find a Maximum Cardinality Pareto Optimal Matching:}}\\
\textbf{Step 1:} \\
Let $G=(A \cup H,E)$ denote the agent-house bipartite graph. \\
Compute a maximum cardinality matching of $G$, denoted as $M$. \\

\textbf{Step 2: }  \\
Let $A'$ denote the set of matched agents in $M$. \\
Create a set of virtual agents $B'$ such that $|A'| + |B'| = |H|$ \\
Let $E'$ denote the edge set \\
Add an edge with weight 0 into $E'$ between each agent in $B'$ and each house in $H$ \\
{\bf forall} $a_i \in A', h_j \in H$ {\bf in parallel}:\\
 \h {\bf if} $M(a_i) = h_j \vee M(a_i) \prec_{a_i} h_j$. \\
   \h\h $E':= E' \cup (a_i, h_j)$; \\
   \h \h$w'(a_i, h_j) := $ rank of $h_j$ in $a_i$'s preference list\\
    {\bf endfor}\\
   $G' = (A' \cup B' \cup H, E', w')$ \\
    Compute a minimum weight perfect matching of $G'$, denoted as $M^{'}$\\
    Output $M^* := \{(a_i,h_j) \in M^{'} ~|~ a_i \in A'\}$ 
\end{minipage}
}
\caption{Algorithm 5: Pareto Optimal Matching for housing allocation\label{fig:house-allocation}}
\vspace{-0.1in}
\end{figure}

 Let $G'(A' \cup B' \cup H, E', w')$ be the graph constructed at Step 2. The following lemma shows the correctness of algorithm 5. 
 \begin{lemma} \label{lem:max-card}
 The matching output by algorithm 5 is a maximum cardinality \textit{Pareto} optimal matching for a housing allocation. 
 \begin{proof}
 Let $M'$ be the minimum weight perfect matching of $G'$. 
 Let $M^{*}$ be the matching output by algorithm 5, which is the induced submatching of $M'$ on the set of matched agents $A'$ after step 1. Step 1 ensures that $M^{*}$ is a maximum cardinality matching. It remains to show that $M^{*}$ is Pareto optimal. Suppose for contradiction that $M^*$ is not Pareto optimal. Then there exists some other matching $M'' \not = M^*$ on real agents such that $M'' \prec M^*$. By definition of Pareto optimality, each agent in $M''$ should be matched to a house at least as good as the house in $M^{*}$ and at least one agent is matched to a strictly better house in $M''$. Since $M^*$ is a maximum cardinality matching, $M''$ must also be a maximum cardinality matching which matches the same set of agents as $M^{*}$. Since the weight of an edge $(a_i, h_j)$ is defined as the rank of $h_j$ at $a_i$'s preference list, we have $w'(M'') < w'(M^{*})$. Since each virtual agent has incident edges of weight 0 in $G'$, there exists another perfect matching formed by edges in $M''$ and some edges incident to virtual agents such that the total weight is smaller than $M^{'}$, contradicting the fact that $M'$ is the minimum perfect matching of $G'$. 
 \end{proof}
 \end{lemma}
 
Now, we can state our main result for the housing allocation problem. 
\begin{theorem}
There is a {\bf RNC}\textsuperscript{2} and quasi-{\bf NC}\textsuperscript{2} algorithm for finding a maximum cardinality \textit{Pareto} optimal matching for the housing allocation problem. 
\begin{proof}
By Lemma \ref{lem:max-card}, the matching obtained by algorithm 5 is a maximum cardinality Pareto optimal matching. The time complexity of algorithm 5 is dominated by the complexity of a minimum weight perfect matching of a graph. By \cite{mulmuley1987matching} and \cite{fenner2016bipartite}, this step can be done in {\bf RNC}\textsuperscript{2} and quasi-{\bf NC}\textsuperscript{2}.
\end{proof}
\end{theorem}

\section{Distributed Algorithms for Housing markets}
In this section, we present a symmetric distributed algorithm to implement the TTC mechanism in a distributed setting. We assume a distributed message passing model with $n$ processes, $p_1, \dots, p_n$, which form a completely connected
topology. The system is synchronous, which means that there is an upper bound on the time for a message to reach its destination. We require that at each round, a node can only send a same message of $O(\log n)$ size to any other node in the network. Since the graph is fully connected, this model is also known as the congest clique model in the literature. Actually, our proposed distributed algorithm fits in a more restricted model called the \textit{broadcast congest clique} model \cite{drucker2014power}, since at round each node only sends the \textit{same} message to all other nodes in the network. This model is in contrast to the \textit{unicast congest clique} model \cite{drucker2014power} which allows each node to send different messages to different nodes in each round.

To implement the top trading cycle algorithm in a fully distributed way, we need efficient distributed algorithms for finding the top trading cycle. Observe that the graph formed by the top choice of each agent is a functional graph since each node has only one outgoing edge. Hence there is only one unique cycle in each connected component of this graph. We present two distributed algorithms for finding all the top trading cycles in a functional graph. 

\subsection{A Las Vegas Algorithm for Finding Cycles in Functional Graphs}
In this section, we give a \textit{Las Vegas} algorithm, shown in Fig. \ref{alg:randomized}, for finding all the disjoint cycles in a functional graph. The primary gradient of the algorithm is a pointer jumping technique. A similar technique is used in \cite{miller1985parallel,vishkin1984randomized} to solve the list ranking problem.

\begin{figure}[htb] 
\begin{centering}
\fbox{\begin{minipage}[t] {3.2in}
\underline{Code for $P_i$:}\\
$active := true$ \\
$succ$: successor of $P_i$ //$P_i$'s next active node\\
$children$: set of nodes that $P_i$ traversed\\
$inCycle$: whether $P_i$ is in the cycle, initially false\\

{\bf while} $active := true$ \\
\h {\bf Coin-flip Step:} \\
\h  Flip a coin, let $myCoin$ denote the result\\
\h Let $succCoin$ be the coin result of $succ$\\
\h{\bf if} $myCoin = head ~\&\&~ succCoin = tail$ \\
\h\h $active:= false$ \\
\h {\bf Explore Step:}\\
\h{\bf if} $active := true$ \\
\h\h Let $succActive$ be the active status of $succ$ \\
\h\h {\bf while} $succActive = false$ \\
 \h  \h\h $children := children \cup {succ}$ \\
 \h\h\h Let $j$ be the successor of $succ$, set $succ := j$ \\
\h\h\h Let $succActive$ be the active status of $succ$ \\
\h\h {\bf endwhile} \\
\h\h {\bf if} $succ = i$ /* now $succ$ is also active /*\\
\h\h\h $active := false$ \\
{\bf endwhile} \\

   \end{minipage}
   \h 
\begin{minipage}[t] {2in}
{\bf Notify Step:}\\ 
{\bf if} $succ = i$ \\
\h Send ("cycle") to $children$ \\
{\bf On receiving} ("cycle"): \\ 
 \h $inCycle := true$ \\
\h Send ("cycle") to all $children$ 
    \end{minipage}}
\end{centering}
\caption{Algorithm 1: Randomized Algorithm for Finding the Cycles \label{alg:randomized}}
\vspace*{-0.1in}
\end{figure}

In this algorithm, each node has a variable \textit{active}, which is initially true. A node terminates the code when \textit{active} becomes false. Each node uses the variable $succ$ to record its current successor node, which initially is its outgoing neighbor. Our algorithm will build a tree. The $children$ variable denotes the current children of a node, which is essentially all the nodes that have been its successor. The $inCycle$ variable denotes whether a node is in the cycle or not. The algorithm is composed of iterations and each node keeps executing an iteration until \textit{active} becomes false. Each iteration includes two steps: a \textit{Coin-flip} step and a \textit{Explore} step. In the \textit{Coin-flip} step, each active node flips a coin. If a node flips head and its successor node flips tail, it becomes inactive. This step is used to reduce the active nodes by a constant fraction. In the \textit{Explore} step, each active node traverses along the path formed by the successor pointer of all nodes and tries to update its successor pointer to be next active node in the path. It also adds all inactive node encountered into its \textit{children} set. When such a active node is found, it checks whether such a node is actually itself, if that is the case, a cycle is detected. After a node determines that it is in the cycle (we will show that there is a unique such node), it broadcasts a cycle message along the tree formed by the child relationship (\textit{Notify} step). We will prove that the set of nodes in the tree rooted at such a node and formed by the child relation is exactly the set of cycle nodes. For the purpose of analysis, we assume the functional graph we consider only has one component, which also means it only has one cycle. Our algorithm works for functional graph with multiple components, since the executions on different components are independent. 

We now show that at the end of algorithm 1 each node correctly knows whether it is in the cycle or not. Let $succ[i]$ denote the value of $succ$ for $P_i$. First, we can easily get the following lemma from the code. 

\begin{lemma} \label{cla:one-cycle-node}
At the end of algorithm 1, there is exactly one node $i$ which has $succ[i] = i$ for each disjoint cycle in the functional graph. 
\begin{proof}
For node $i$, if $succ[i] = j$ at some point, then there exists a directed path from node $i$ to node $j$. 
Let us consider a single connected component of the functional graph. 
For any non-cycle node, its $succ$ cannot be itself since it does have a directed path to itself. Hence, it is sufficient to consider only cycle nodes. 
We first show there is at least one node $i$ with $succ[i] = i$. We claim that there is at least one active cycle node remaining after the \textit{Coin-flip} step of each iteration. To become inactive, a node has to flip head and its successor has to flip tail. This implies that two consecutive active nodes cannot become inactive simultaneously. Hence, at the end of the algorithm at least one cycle node $i$ with $succ[i] = i$. Also, it is obvious that at most one cycle node $i$ can have $succ[i] = i$ at the end of the algorithm. 
Therefore, there is exactly one node $i$ which has $succ[i] = i$ for each disjoint cycle in the functional graph.
\end{proof}
\end{lemma}
Let $i$ be the node with $succ[i] = i$ at the end of the algorithm. Let $T$ be the tree rooted at node $i$ and constructed from the \textit{child} relation at the end of the algorithm. Let $V_T$ denote the set of nodes in tree $T$. Let $C$ denote the set of nodes in the cycle. 

\begin{lemma} \label{lem:cycle_tree}
$C=V_T$
\begin{proof}
We prove $C \subseteq V_T$ and $V_T \subseteq C$.
Suppose node $i$ runs for $L$ iterations. Let $A_r$ denote the set of active cycle nodes at round $r$, $1 \leq r \leq L$. To prove $C \subseteq V_T$, we show by induction that each node in $A_r$ is in tree $T$ for all $r$. 

Base case, $r = L$. $A_L = \{i\}$. Node $i$ is the root of $T$. 

Induction case: Suppose each node in $A_k$ is in $T$. We need to show that each node in $A_{k - 1}$ is in $T$. It is sufficient to show that the nodes in $A_{k - 1}$ which become inactive at round $k$ are in $T$. Since all active cycle nodes at each round still form a cycle, $A_k$ divides $A_{k - 1}$ into multiple directed paths. For any path $P$ of form $v_i,v_{i + 1},...,v_{j}$, only the two end nodes $v_i$ and $v_j$ are in $A_k$. From the code we know that node $v_i$ continues to find active nodes along $P$ at round $k$, and it stops until it reaches node $v_j$. Thus, all nodes in path $P$ between $v_i$ and $v_j$ become the children of node $v_i$. So, all nodes in path $P$ are in $T$. Hence, all nodes in $A_{k - 1}$ are in $T$. Therefore, we have $A_r$ is in $T$ for any $1 \leq r \leq L$. Since $A_1$ is exactly the set of cycles nodes, we have all cycles nodes are in $T$. Thus, $C \subseteq V_T$. 

To prove $V_T \subseteq C$, we show that the any non cycle node is not in tree $T$. For any non cycle node $j$, suppose $j \in V_T$. Then $j$ must be a descendant of root node $i$. From the algorithm we know that the children relation is formed by next relation in the original graph. Thus, there must be a directed path from node $i$ to node $j$ in the original graph. This means $j$ must be in the unique cycle, a contradiction. 
\end{proof}
\end{lemma}

\begin{theorem}
Algorithm 1 computes all the cycles of a functional graph $G$. It has round complexity of $O(\log l)$ and message complexity of $O(n\log l)$, w.h.p, where $l$ is the length of the longest cycle in $G$. 
\begin{proof}
Since the $cycle$ message only traverses through tree $T$, from Lemma \ref{lem:cycle_tree}, we know that each cycle node receives the $cycle$ message and each non cycle node does not receive the $cycle$ message.

Since the number of active nodes in any cycle reduces by a constant fraction in expectation at each iteration and each iteration takes constant number of rounds, by Chernoff bound, algorithm 1 takes $O(\log l)$ rounds w.h.p. 
Each round of the algorithm takes at most $O(n)$ messages, which results in $O(n \log l)$ messages in total.
\end{proof}
\end{theorem}

\subsection{A Deterministic Algorithm for Finding Cycles in Functional Graphs}
In this section, we present a deterministic algorithm for finding all the disjoint cycles in a functional graph, shown in Fig. \ref{alg:deterministic}. This algorithm is similar to the \textit{las vegas} algorithm in the previous section, with only one key difference. We replace the \textit{Coin-flip} step in algorithm 1 to the \textit{Coloring} step. Observe that in algorithm 1 the primary purpose of the \textit{Coin-flip} step is to reduce the number of active nodes by a constant factor while ensuring that any two consecutive active cycle nodes cannot become inactive at the same time. Graph coloring techniques can also serve this purpose. Hence, we simply replace the \textit{coin-flip} step in algorithm 1 to be a \textit{Coloring} step, which is an invocation of the 6-coloring algorithm due to \cite{cole1986deterministic}. After the \textit{Coloring} step, each node compares its color with the color of its successor. If a node has a smaller color than its successor, it becomes inactive. Then all remaining active nodes perform the \textit{Explore} step as in algorithm 1. 

\begin{figure}[htbp] 
\begin{centering}
\fbox{\begin{minipage}[t]  {5in}
\underline{Code for $P_i$:}\\
/* Variables are the same as algorithm 1 */\\
{\bf while} $active := true$ \\
\h {\bf Coloring Step:} \\
\h 6-coloring of active nodes using coloring algorithm from \cite{cole1986deterministic} \\
\h Request the color of $succ$, denoted as $c'$ \\
\h{\bf if} $c < c'$ /* If my color is less than the color of my successor, becomes inactive */\\
\h\h $active:= false$ \\
\h Execute {\bf Explore Step} of {\bf Algorithm 1} \\
{\bf endwhile}
\end{minipage}
} 
\end{centering}
\caption{Algorithm 2: Deterministic Algorithm for Finding Cycles  \label{alg:deterministic}}
\vspace*{-0.1in}
\end{figure}

We can observe that after the coloring step at each iteration, the node with the largest color remains active in each disjoint cycle. By similar argument, we can show that Lemma \ref{cla:one-cycle-node} and Lemma \ref{lem:cycle_tree} still hold. 

\begin{theorem}
Algorithm 2 computes all the disjoint cycles in a functional graph and takes $O(\log^* n \log l)$ rounds and $O(n\log ^*n \log l)$ messages. 
\begin{proof}
Since no more than 5 consecutive active nodes become inactive at each iteration by the property of 6-coloring, the \textit{Explore Step} still takes constant rounds. The coloring step introduces an additional $O(\log^*n)$ factor. Thus, algorithm 2 terminates in $O(\log^* n \log l)$ rounds and takes $O(n\log^*n \log l)$ messages.
\end{proof}
\end{theorem}

\subsection{Distributed Top Trading Cycle Algorithm}
We now present a distributed version of the top trading cycle algorithm. As in the sequential setting, we assume that each node knows which nodes are holding the houses in its preference list. Indeed, every node can broadcast its house to all. This only requires one round and $O(n^2)$ messages.

\begin{figure}[htbp] 
\begin{centering}
\fbox{\begin{minipage}  {2.6in}
\underline{Code for $P_i$:}\\
/* Variables */ \\ 
$next_i$: the node which holds current top choice of $P_i$ \\
$assigned_i$: whether be assigned final house \\ 
$h_i$: the house $P_i$ holds \\
$succ_i$: successor of $P_i$, same as algorithm 1 \\
$pref_i$: mapping from a house to the node which holds the house. \\

\textbf{One Stage:} \\
\h $succ_i = next_i$\\
\h Execute {\bf Algorithm 1 or 2} to find out \h cycle nodes \\
\h {\bf if $P_i$} in cycle \\
\h \h Let $h_j$ denote the house of $next_i$\\
\h\h $h_i := h_j$ \\
\h \h $assigned := true$ \\
\h \h Broadcast $remove(h_i)$ to all \\
\h \h {\bf if} $P_i$ has no children \\
\h \h\h Send $ok$ to its parent \\
\end{minipage}
\h
\begin{minipage}  {2.6in}
{\bf On receiving} $ok$ from all children: \\
\h {\bf if} $P_i$ is the root of the tree \\
\h \h Broadcast $nextStage$ to all \\
\h {\bf else} \\
\h\h Send $ok$ to its parent \\

{\bf On receiving} $nextStage$: \\
\h {\bf if} $assigned_i := false$ \\
\h\h $active := true$ \\
\h\h Let $Top_i$ denote the next available top choice \\
 \h \h $next_i := pref_i[Top_i]$ \\
\h\h Start next stage \\

{\bf On receiving} $remove(h_j)$ from $j$: \\
 \h Remove $h_j$ from $pref_i$ 
\end{minipage}}
\end{centering}
\caption{Algorithm 3: Distributed Version of Top Trading Cycle\label{alg:house}}
\end{figure}

The distributed algorithm is shown in Fig. \ref{alg:house}. The basic idea is using the cycle finding algorithms presented above to simulate each stage of the top trading cycle algorithm. During a stage, all nodes first build the top choice functional graph, i.e, update their $succ$ variable to be the node which holds their current top choice. Then, all nodes execute algorithm 1 or 2 to find out whether they are in a cycle or not. After that, a cycle node gets assigned its current top choice and broadcasts a $remove$ message which contains the assigned house to all nodes. When node $P_i$ receives $remove$ messages from other nodes, it deletes the houses contained in the messages from the preference list, i.e, from $pref_i$.  When executing algorithm 1 or 2, nodes might terminate at different rounds. Thus, we need to coordinate the execution of each stage. In order to achieve this, we use a convergecast step using the tree built in the execution of algorithm 1 or 2. When a node completes broadcasting its $remove$ message to all, it sends a $ok$ message to its parent in the tree if it is a leaf node in the tree. For non-leaf nodes, they send an $ok$ message to their parents only when they receive $ok$ messages from all children. For the root node, when it receives $ok$ from all its children, which means all nodes have updated their preference list, it broadcasts a $nextStage$ message to all to notify all nodes to start the next stage of the algorithm.

Since each stage of algorithm 3 simulates each iteration of the TTC mechanism. The correctness of algorithm 3 follows from the correctness of TTC. We now look at the round and message complexity. 
\begin{theorem}
Algorithm 3 computes the core of a housing market in $O(n)$ rounds and takes $O(n^2)$ messages. 
\begin{proof}
We just analyze the complexity of adopting the {\em Las Vagas} algorithm as a subroutine. The complexity of the deterministic algorithm just has an additional $\log^*n$ factor. Let $l_i$ denote the length of the cycle at stage $i$ of algorithm 3. At stage $i$, both finding the cycle and convergecast along the tree need $O(\log l_i)$ rounds, w.h.p. Finding the cycle takes $O(n \log l_i)$ messages and convergecast takes $O(l_i)$ messages. Thus, each stage takes  $O(\log l_i)$ randomized rounds and $O(n \log l_i)$ messages. Therefore, since $\sum {l_i} = n$, algorithm 3 takes $O(n)$ rounds and $O(n^2)$ messages in the worst case. 
\end{proof}
\end{theorem}

\section{Conclusion} 
We conclude with two open problems. We have shown that computing a Pareto optimal matching for a housing allocation is in {\bf CC}, which yields a linear time and linear work parallel algorithm. Computing an individual Pareto optimal matching for a housing market seems harder. It is interesting to know the relationship between this problem and the {\bf CC} class. It is unlikely to be {\bf CC}-complete, since this would imply a {\bf RNC}\textsuperscript{2} and a quasi-{\bf NC}\textsuperscript{2} algorithm for the {\bf CC} class. We also show that computing the core of a housing market is {\bf CC}-hard by giving a logspace reduction from the LFMM problem. It is interesting to know whether this problem is {\bf CC}-complete. Or can we show that it is {\bf P}-complete?

\bibliography{ref}

\begin{thebibliography}{10}

\bibitem{abdulkadirouglu1998random}
Atila Abdulkadiro{\u{g}}lu and Tayfun S{\"o}nmez.
\newblock Random serial dictatorship and the core from random endowments in
  house allocation problems.
\newblock {\em Econometrica}, 66(3):689--701, 1998.

\bibitem{abdulkadirouglu1999house}
Atila Abdulkadiro{\u{g}}lu and Tayfun S{\"o}nmez.
\newblock House allocation with existing tenants.
\newblock {\em Journal of Economic Theory}, 88(2):233--260, 1999.

\bibitem{abraham2004pareto}
David~J Abraham, Katar{\'\i}na Cechl{\'a}rov{\'a}, David~F Manlove, and Kurt
  Mehlhorn.
\newblock Pareto optimality in house allocation problems.
\newblock In {\em International Symposium on Algorithms and Computation}, pages
  3--15. Springer, 2004.

\bibitem{amira2010distributed}
Nir Amira, Ran Giladi, and Zvi Lotker.
\newblock Distributed weighted stable marriage problem.
\newblock In {\em International Colloquium on Structural Information and
  Communication Complexity}, pages 29--40. Springer, 2010.

\bibitem{cole1986deterministic}
Richard Cole and Uzi Vishkin.
\newblock Deterministic coin tossing with applications to optimal parallel list
  ranking.
\newblock {\em Information and Control}, 70(1):32--53, 1986.

\bibitem{cook2014complexity}
Stephen~A Cook, Yuval Filmus, and Dai Tri~Man Le.
\newblock The complexity of the comparator circuit value problem.
\newblock {\em ACM Transactions on Computation Theory (TOCT)}, 6(4):15, 2014.

\bibitem{david2013algorithmics}
Manlove David.
\newblock {\em Algorithmics of matching under preferences}, volume~2.
\newblock World Scientific, 2013.

\bibitem{drucker2014power}
Andrew Drucker, Fabian Kuhn, and Rotem Oshman.
\newblock On the power of the congested clique model.
\newblock In {\em Proceedings of the 2014 ACM symposium on Principles of
  distributed computing}, pages 367--376. ACM, 2014.

\bibitem{fenner2016bipartite}
Stephen Fenner, Rohit Gurjar, and Thomas Thierauf.
\newblock Bipartite perfect matching is in quasi-nc.
\newblock In {\em Proceedings of the forty-eighth annual ACM symposium on
  Theory of Computing}, pages 754--763. ACM, 2016.

\bibitem{hylland1979efficient}
Aanund Hylland and Richard Zeckhauser.
\newblock The efficient allocation of individuals to positions.
\newblock {\em Journal of Political economy}, 87(2):293--314, 1979.

\bibitem{karp1988survey}
Richard~M Karp.
\newblock A survey of parallel algorithms for shared-memory machines.
\newblock 1988.

\bibitem{ma1994strategy}
Jinpeng Ma.
\newblock Strategy-proofness and the strict core in a market with
  indivisibilities.
\newblock {\em International Journal of Game Theory}, 23(1):75--83, 1994.

\bibitem{maggs2015algorithmic}
Bruce~M Maggs and Ramesh~K Sitaraman.
\newblock Algorithmic nuggets in content delivery.
\newblock {\em ACM SIGCOMM Computer Communication Review}, 45(3):52--66, 2015.

\bibitem{mayr1992complexity}
Ernst~W Mayr and Ashok Subramanian.
\newblock The complexity of circuit value and network stability.
\newblock {\em Journal of Computer and System Sciences}, 44(2):302--323, 1992.

\bibitem{miller1985parallel}
Gary~L Miller and John~H Reif.
\newblock Parallel tree contraction and its application.
\newblock Technical report, HARVARD UNIV CAMBRIDGE MA AIKEN COMPUTATION LAB,
  1985.

\bibitem{mulmuley1987matching}
Ketan Mulmuley, Umesh~V Vazirani, and Vijay~V Vazirani.
\newblock Matching is as easy as matrix inversion.
\newblock In {\em Proceedings of the nineteenth annual ACM symposium on Theory
  of computing}, pages 345--354. ACM, 1987.

\bibitem{roth1982incentive}
Alvin~E Roth.
\newblock Incentive compatibility in a market with indivisible goods.
\newblock {\em Economics letters}, 9(2):127--132, 1982.

\bibitem{roth1977weak}
Alvin~E Roth and Andrew Postlewaite.
\newblock Weak versus strong domination in a market with indivisible goods.
\newblock {\em Journal of Mathematical Economics}, 4(2):131--137, 1977.

\bibitem{shapley1974cores}
Lloyd Shapley and Herbert Scarf.
\newblock On cores and indivisibility.
\newblock {\em Journal of mathematical economics}, 1(1):23--37, 1974.

\bibitem{vishkin1984randomized}
Uzi Vishkin.
\newblock Randomized speed-ups in parallel computation.
\newblock In {\em Proceedings of the sixteenth annual ACM symposium on Theory
  of computing}, pages 230--239. ACM, 1984.

\bibitem{yuan1996residence}
Yufei Yuan.
\newblock Residence exchange wanted: a stable residence exchange problem.
\newblock {\em European Journal of Operational Research}, 90(3):536--546, 1996.

\bibitem{zhou1990conjecture}
Lin Zhou.
\newblock On a conjecture by gale about one-sided matching problems.
\newblock {\em Journal of Economic Theory}, 52(1):123--135, 1990.

\end{thebibliography}

\end{document}